\documentclass[a4paper]{jpconf}
\usepackage{graphicx}
\usepackage{amsmath}
\usepackage{amssymb}
\usepackage{mathtools}
\usepackage{bm}
\usepackage{citesort}
\begin{document}
%
%
\title{Pulsar rotation with superfluid entrainment}
\author{
 Marco Antonelli, 
    Pierre M. Pizzochero
    }
\address{
    Department of Physics, 
    Universit\`a degli Studi di Milano, 
    and Istituto Nazionale di Fisica Nucleare, 
    Via Celoria 16, 
    20133 Milano, 
    Italy
    }
\ead{
    marco.antonelli@unimi.it
    }
\begin{abstract}
%
Large pulsar glitches (like the ones detected in the Vela) are though to be a consequence of the superfluid component present in the interior of mature neutron stars:
this component can rotate differentially with respect to the normal part of the star, storing the angular momentum needed to produce the observed sudden decrease
of the pulsar rotational period.
However strong entrainment (a non-dissipative effect that couples the superfluid component with the non-superfluid component inside the star) 
challenges this picture.
Here we study the impact of entrainment on the angular momentum that can be exchanged between the normal component and the superfluid 
during a glitch by means of a consistent 
global model. 
This allows to estimate the maximum angular momentum reservoir stored into the superfluid component of the star: 
the essential ingredient are newly calculated mesoscopic pinning forces that block the superfluid vorticity in the crust of the neutron star.
This method can also provide a quantitative test for global models of rotating neutron stars, as well as for microphysical inputs
present in literature (like entrainment parameters and pinning forces).
\end{abstract}
\section{Introduction}
\vspace{2mm}
Glitches  are sudden jumps of amplitude $\Delta\Omega$ in the angular velocity $\Omega$ of some isolated pulsars
with fractional increases that span the range $\Delta\Omega/\Omega \sim 10^{-11} \div 10^{-5}$ 
in otherwise steadily spinning-down neutron stars (NSs).
According to the Jodrell Bank Glitch Catalog\footnote{Glitch data are extracted from the database www.jb.man.ac.uk/pulsar/glitches.html.},
more than four hundred glitches have been observed, in more than a hundred objects including rotational-powered pulsars,
magnetars and millisecond pulsars.
%
Although there is some evidence that small glitches in the young Crab pulsar are caused by star quakes,
the widely accepted explanation for large glitches is based on the presence of a superfluid component
(made of paired neutrons) that can rotate differentially with respect to the non-superfluid rest of the NS.
The superfluid stores extra angular momentum 
that can be abruptly exchanged with the normal component, resulting in a little decrease of the observed rotational period of the pulsar
(for a review on glitch models see \cite{HM15}). 
A rotating superfluid is permeated by many vortex lines (of the order $\sim 10^{19}$ for the Vela pulsar) of quantized circulation.
The exchange of angular momentum is mediated by these vortices \cite{ANDERSON1975}  
that
induce a  
mutual friction between the superfluid and normal components \cite{Andersson2006}.
%
The two physical ingredients that currently challenge the description of pulsar glitches, both at the mesoscopic scale of vortex lines as well as at the macroscopic stellar scale,  
are entrainment and pinning. 
%
\\
\\
{\it Entrainment - }
A key feature of superfluidity is the possibility for the normal and superfluid components to 
flow independently although they compenetrate. 
The formalization of this multi-fluid problem in the interior of NSs is based on the seminal works on the hydrodynamics of He-II 
that extended the early two-fluid models of Tisza and Landau \cite{HV56,BK61}. 
Later, with the aim to study superfluid solutions, Andreev and Bashkin developed a three-fluid framework where two superflows and a normal flow exist simultaneously \cite{AB75}. 
They found that entrainment (a non-dissipative effect that couples the two species)
can be described in the frame of reference comoving with the normal flow
as a ``mass-matrix'' which relates the momentum of one constituent to the 
kinematic velocities  of both constituents (see also \cite{AnderssonReview,chamelReview,RZbook} for modern formalism 
with applications to the hydrodynamics of NS interiors). 
Thus entrainment is also discussed, both in  the core and in crust, in terms of effective masses (which definition depend on the reference frame used).
Dripped neutrons in NS crusts are analogous to electrons in ordinary metals and their properties can be
determined by the band theory of solids \cite{C12}: only neutrons in the conduction band can move throughout the crust.
Despite the absence of viscous drag, the crust resists the neutron current due to Bragg scattering.
This is a non-local and non-dissipative effect: neutrons are scattered by individual clusters and can interfere constructively or
destructively. In the latter case neutrons cannot propagate, as if they were very massive \cite{CC05}.
The scenario is different in the core, where neutrons and protons can only be locally scattered by surrounding nucleons \cite{CH06}.
The resulting low effective mass of neutrons in the core can be interpreted in terms of a back-flow of nucleons. This local effect also exists 
in the crust but it is negligible when compared to Bragg scattering. 
\\
Entrainment challenges our understanding of pulsar glitches. 
In particular Vela glitches have been thought to arise from the superfluid in neutron-star crust \cite{Alpar1984a,LE99}.
However superfluid can be strongly entrained by the crust and this results
in a decreased mobility of the crustal superfluid that cannot carry enough angular momentum to explain large glitches \cite{AG12,C13}.
In our numerical examples we use the effective neutron masses 
given in \cite{CH06} and in \cite{C12} for the core and the crust respectively. 
\\
\\
{\it Pinning - }
Pinning is the mechanism that allows to build up the angular momentum reservoir \cite{ANDERSON1975}: when vortices are blocked
the normal component slows down due to radiation losses while the pinned superfluid maintains its state of motion, 
lagging behind and storing angular momentum which can then be released in a glitch.
\\
Pinning is a well known phenomenon that occurs in type-II superconductors, where the flux-tubes can pin to defects of the crystalline 
structure.
%
%
%
Something analogous happens in NSs:
when vortex lines are immersed into a background  current (of superfluid neutrons) an hydrodynamical lift develops and the vortices tend 
to drift outside the star (vortex-creep). Strong pinning allows to support higher currents before the dissipative effects generated by the motion of vortices
enter the game; this in turns imply that a bigger angular momentum reservoir can be stored into the current of superfluid neutrons.
The strength of pinning is described by a mesoscopic quantity $f_P$ that defines the depinning threshold:
it is the module of the maximum lift force on vortices that the pinning centers can sustain.
At a microscopic scale it is possible to calculate the {\it individual pinning energies}
of a single vortex with a single pinning center \cite{DP06} but to find the maximal non-dissipative current of neutrons that pinning can sustain
we need the {\it average pinning force} of vortices passing trough many pinning sites \cite{SP15}.
Typically in NSs pinning refers to the interaction of vortices with the crustal lattice and pinning centers are the nuclei of the inner-crust.
In this work we will not consider the further possibility of pinning between vortex lines and flux-tubes of the proton superconductor,
even if this suggests interesting scenarios \cite{GA14}. However a vortex-fluxtube pinning profile, that is a consequence of entrainment in the core, can be added to
that of crustal pinning we use, without any change in our model.
%
%
\\
\\
In this presentation we expand the discussion of the model proposed in  \cite{AP16}. 
This model, despite its simplicity, can be used to test quantitatively 
the widespread assumption of a stable vortex array in superfluid NSs.
In fact, vortex-creep models (since the seminal work of \cite{Alpar1984II}) do not account for entrainment or stratification.  
Similarly, global models based on the two-fluid formalism have either neglected entrainment \cite{HP12} or have been studied under 
the assumption of uniform entrainment and rigid rotation of the neutron superfluid \cite{PC02,SP10}.
\section{Application to pulsar glitches: axially symmetric model}
\vspace{2mm}
The multi-fluid formalism used to describe the superfluid hydrodynamics in NS interiors is well established and in principle enables 
us to model the glitch and the subsequent relaxation \cite{PM06,HP12,HH16,SC16}. 
However in a three-dimensional hydrodynamical simulation it is not simple to take into account consistently for 
stratification, pinning forces and entrainment. 
We show how to implement consistently these ingredients within a simplified geometry, 
as a first natural approximation to the full problem.
In this way we rule out a priori turbulent motion and its consequences on the vortex-mediated mutual friction \cite{AS07,PM06}: 
this is a limit of the model to be kept in mind.
The model is already detailed in \cite{AP16}, in this section we discuss its features and assumptions: 
\begin{itemize}
    \item The star spins around the z-axis 
    \footnote{We use cylindrical coordinates $(x,\theta, z)$. The spherical radius $r$ should be read as $\sqrt{x^2+z^2}$.} 
    (there is no precession); since we study non-millisecond pulsars, we consider that the equilibrium structure is given by solving the TOV equations.
    As sketched in Fig 1, the superfluid is contained into a spherical volume of radius $R_d$ 
    (the radius at which the neutron drip starts, corresponding to a density of $\sim 4\times10^{11}\,$g cm$^{-3}$).
    \item In the absence of a layer of normal matter between the core and the crust \cite{ZS04},
    the vortex cores can pass continuously through this smooth transition.
    Leaving out the problem of the unknown composition of the inner core, we consider that the superfluid component 
    (the n-component) is threaded by straight vortices parallel to $\hat{\bm{e}}_z$. 
    However it is equally simple to consider superfluidity only inside a spherical shell of given radii.
    The scenario of continuous vortex lines was already suggested in \cite{R76} as an alternative to the scenario of distinct vorticity in the crust and in the core. 
    \item The rigid crust and the charged component (i.e. electrons, protons and bounded neutrons, that we call p-component)
    are locked into the strong magnetic field of the neutron star on short timescales  \cite{E79}. 
    Electromagnetic losses produce a constant braking torque $-I \dot\Omega_{\infty}{\bm{e}}_z$ that acts on the p-component, 
    where $I$ is the total moment of inertia of the star and $\dot\Omega_{\infty}$ the absolute value of the spin-down rate over a long period of time.
    The velocity field of this rigid p-component is $\bm{v}_p = x \Omega_p \hat{\bm{e}}_\theta$, which gives the observed pulsar period $2\pi \Omega_p^{-1}$. 
    \item The presence of vortices induces a mutual friction between the n-component and the p-component: 
    scattering of particles (or quasiparticles) off vortex cores gives rise to mesoscopic drag force $\bm f_D$ \cite{Alpar1984a,J90a,EB92,Andersson2006}.
    \item Vortex lines can pin to nuclei in the solid inner crust. The force per unit length $f_p(\rho)$ expresses the strength of the vortex-lattice 
    interaction \cite{SP15}. 
\end{itemize}
We are thus far from describing the realistic hydrodynamical problem of a NS interior; 
superfluid turbulence (a well known phenomenon in He-II where it gives rise to a complex tangle of vortex lines \cite{Qbook}) is likely
to  develop in some regime also in NSs \cite{PM06}, but its effect on macroscopic hydrodynamics and its relevance for glitch models are still debated.
In particular the tension of a single line is much smaller than 
the hydrodynamical force acting on it, which suggests bending of vortices, such that the core is susceptible to become turbulent \cite{AS07}.
However this argument does not account for the stabilizing effect of rotation: the increased effective tension of a bundle of 
vortices as compared to a single vortex line resists bending and the array of vortex lines could remain parallel to the rotation axis.
This can be seen as a generalization of the Taylor-Proudman theorem to axisymmetric and spinnig-down frictionless fluids \cite{RS74}. 
\begin{figure}[h]
    \includegraphics[width=18pc]{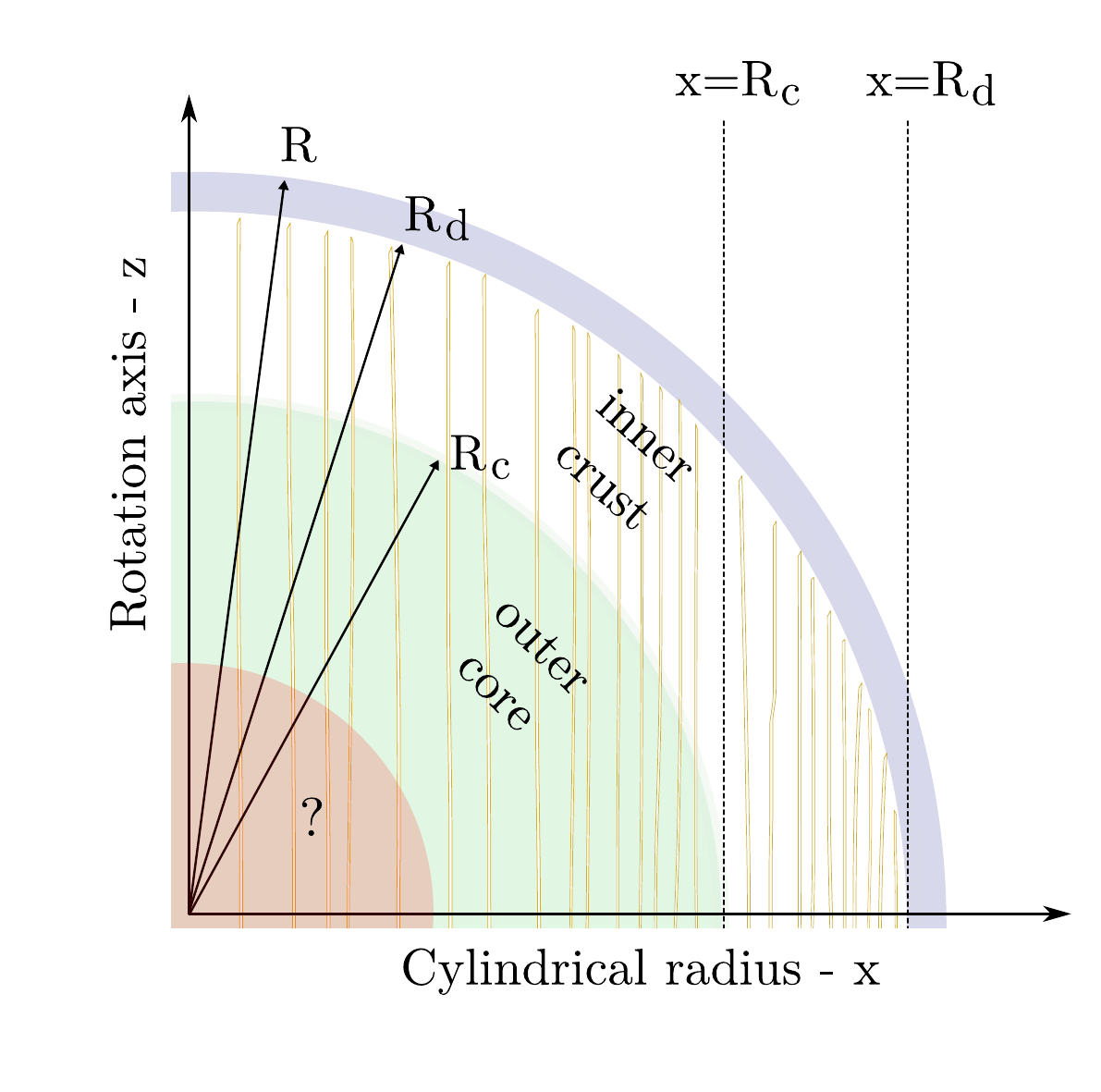}
    \hspace{1pc}
        \begin{minipage}[b]{17.5pc}
            \caption{
                Sketch of the stellar structure (out of scale), with the geometrical definitions used. In the cylindrical shell
                $R_c<x<R_d$ the vortex lines are completely immersed in the inner crust. The outer crust (the thin blue layer in the figure)
                is part of the p-component. In principle vortices can stop at any spherical radius and all the input quantities of the model 
                are consistently adjusted by following the prescriptions detailed in ref. \cite{AP16}.
                \\            
                \\
            }
            \label{fig:structure}
    \end{minipage}
\end{figure}
%
%
\\
In the zero temperature limit the microphysical quantities of interest, as the the superfluid fraction $x_n(\rho)$ and the dimensionless effective mass 
$m^*(\rho)= 1-\epsilon_n(\rho)$ of entrained neutrons, can be easily parametrized as functions of the spherical radius $r$ after solving the TOV equations for a given EOS.
We impose a Newtonian dynamics on the general relativistic stellar structure, thus the total moment of inertia is given by  
\[
I \, = \, \frac{8 \pi}{3}\int_0^R \! r^4 \rho(r) \,\,dr \, . 
\]
General relativistic corrections to this quantity can be computed using the Hartle slow-rotation approximation, 
where the background structure is still given by the TOV equations and are expected to be large (see e.g. \cite{RZbook} for a recent description of this approximation). 
We further discuss this point in the concluding remarks.
\\
\\
{\it Conservation of vorticity - } 
Following the Newtonian formalism outlined in \cite{PC02}, the momentum per particle of the superfluid is
a linear superposition of the local kinematic velocities of both components:
%
\[
\bm{p}_n/m_n \,=\,
\left(1-\epsilon_n\right) \bm{v}_n\,
+\,\epsilon_n \bm{v}_p \, ,
\]
where $m_n$ is the neutron bare mass. At the microscopic scale $\bm{p}_n$ is irrotational but here we consider it as averaged over a region that contains 
many vortex lines. 
In our axially symmetric configuration the azimuthal component of the momentum turns out to be 
columnar (it has no dependence on $z$). It is thus convenient to introduce an auxiliary angular velocity $\Omega_v(x)$ defined by
\begin{equation*}
\bm{p}_n(\bm{x})\cdot \hat{\bm{e}}_\theta = m_n\,x \,\Omega_v(x)  \,\, .
\end{equation*}
On the other hand, the angular velocity of the n-component is non-columnar due to the (spherically symmetric) density dependence of the 
entrainment parameter $\epsilon_n(r)$.
Exchange of angular momentum between the superfluid and the p-component is possible only if we allow for radial motion of vortices.
Let $v_L^x$ be the radial component of the local mean velocity of vortex lines (the velocity of a vortex is normal to the line),
then a simple argument early proposed in \cite{Alpar1984II} gives 
\begin{equation}
\label{eq:1}
\dot\Omega_v(x,t) = -\left( 2\Omega_v(x,t) + x\partial_x \Omega_v(x,t) \right)\frac{v_L^x(x,t)}{x} \,\, .
\end{equation}
This equation expresses a conservation law for vortex lines flowing trough a cylindrical shell of radius $x$. 
The radial velocity $v_L^x(x,t)$ needs to be expressed in terms of the variables $ \Omega_v(x,t)$ and $\Omega_p(t)$
by solving an explicit model of vortex dynamics. In \cite{AP16} a simple example is proposed: we construct 
a particular form of the friction functional $\mathcal{B}$ defined as
\[
v_L^x(x) \,\, = \,\, x \,\, \mathcal{B}[\Omega_v,\Omega_p,x] \,\, (\Omega_v(x)-\Omega_p) \,. 
\]
The explicit form of $\mathcal{B}$ depends on how vortex dynamics is modeled; correlated motion of many lines, pinning and
finite tension of macroscopically long vortices make the explicit construction of this functional a non trivial task. 
\\
\\
{\it Total angular momentum balance - } The isolated pulsar dynamics is slowly driven by angular momentum losses due to radiation emission, namely
\[
\int\!d^3x\,\rho(r)\,\left[ \, 
    x_n(r) \, \bm{x} \times \dot{\bm{v}}_n 
    + 
    (1-x_n(r)) \, x^2\, \dot{\Omega}_p\, \hat{\bm{e}}_z \,
\right] \,=\,-I\,\dot\Omega_{\infty}\hat{\bm{e}}_z   \, .
\]
Expressing the non-columnar velocity of the n-component in terms of the angular velocities,
the angular momentum balance takes the form  
\begin{equation}
\label{eq:2}
I_p \, \dot{\Omega}_p \, + \, I_v \, \langle \, \dot{\Omega}_v \, \rangle \, = \, -I \, \dot{\Omega}_\infty \, ,
\end{equation}
where $\Omega_v$ is averaged over the cylindrical radius $x$, namely 
\[
    \langle \, \dot{\Omega}_v \, \rangle \, = \, \frac{1}{I_v} \int_0^{R_d} \! dI_v(x) \,\, \dot{\Omega}_v(x,t) \, .
\]
The partial moments of inertia are $I_p = I-I_v$ and $I_v$, that is also the normalization factor of the measure $dI_v$.
The physical constraint on entrainment parameter $\epsilon_n < 1-x_n$  \cite{CH06} guarantees the positivity of this quantity,
namely $I_v<I$.
The explicit Newtonian form of $dI_v$ is
\[
    \frac{dI_v(x)}{dx} \,=\, 4\,\pi\,x^3  \int_{0}^{z(x)} \! dz \,\,  \frac{\rho_n(r)}{m^*(r)} \, ,
\]
with $z(x)$ the height of vortices passing through $x$.
The  appearance of the effective mass at the denominator is physically reasonable: the effect of entrainment 
is to redefine the mass of neutrons (i.e. $\bm p_n = m^*\bm v_n$ in the frame corotating with the p-component) that rescales $\rho_n$.
\\
At the steady-state (i.e when $\dot \Omega_p=\dot \Omega_v$ for all the $x$ values ),
Eq \ref{eq:2} assures that the star responds to the braking torque as a whole with a spin-down rate of $-\dot{\Omega}_\infty$. 
Thus, provided that the braking torque itself is not evolving, the value of $\dot{\Omega}_\infty$ coincides with  the absolute value of the 
observed mean spin-down rate over a long period of time (even if in this period the pulsar glitched several times).
It is also useful to introduce the lag $\omega$, defined as
\[
\omega(x,t) \, = \, \Omega_v(x,t) - \Omega_p (t) \, .
\] 
At the steady state the lag is constant in time and strictly positive for all the $x$ values.
\\
\\
{\it Friction functional - } 
Vortex lines experience a viscous drag force 
per unit length $\bm{f}_D$ and an hydrodynamical lift (the Magnus force) $\bm{f}_M$ expressed by 
\begin{align*}
\bm{f}_D &= - \eta(r) ( \bm{v}_L(\bm{x})-\bm{v}_p(\bm{x}) ) \, ,
\\
\bm{f}_M &=\rho_n(r) \bm{\kappa}(\bm{x}) \times ( \bm{v}_L(\bm{x})-\bm{v}_n(\bm{x}) ) \, .
\end{align*}
As discussed in \cite{Andersson2006} the vector $\bm{\kappa}$ is given by
\[
\bm{\kappa}(\bm{x}) \,=\, \kappa \,\nabla \times\bm{p_n} \, / \, |\nabla \times \bm{p_n}| \, .
\]
Within our working assumptions, this field is constant and parallel to  $\hat{\bm{e}}_z$.
To implement this infinite rigidity of vortex lines, the equation of motion of vortices is modeled as
\[
\bm{F}_{tot} \,=\, \int_L \! dl(\bm{x}) \,[ \bm{f}_M(\bm{x})+\bm{f}_D(\bm{x}) ] \,=\,0 \, ,
\]
where the integration is performed over the vortex line. 
Solving the above equation for the radial component of the vortex velocity gives
that the friction functional $\mathcal{B}$ is just a function $\mathcal{B}^x$ of the cylindrical radius 
\begin{equation}
    \label{eq:free-velocity}
    v_L^x \, = \, x \, \mathcal{B}^x(x) \,\omega(x,t) \, .
\end{equation}
At this level we still have to account for pinning: vortex lines that are perfectly pinned are blocked 
and forced to corotate with the p-component, until the Magnus force induced to vortex lines by the background flow of neutrons overcomes the depinning critical value.
This threshold effect introduces explicit and non-smooth dependence of $\mathcal B$ on the dynamical variables.
The depinning condition can be found by imposing that
\[
    |\bm{F}_{tot}|_{\bm v_L =\bm v_p} \,=\, \int_L \! dl \,\frac{\kappa\,\rho_n}{m*} \,x\,\omega  \,=\, \int_L \! dl \,f_P
\]
and a straightforward calculation gives that the second equality is realized when the dynamical lag $\omega$ reaches the value 
\begin{equation}
\label{eq:clag}
\omega_{cr}(x) \, = \, \frac{ \int_{0}^{z(x)} \! dz \,\,  f_P(r) }{\kappa \, x\,  \int_{0}^{z(x)} \! dz \,\,\rho_n(r)/m^*(r)  } \, .
\end{equation}
This function of the cylindrical radius (sketched in Fig 2 for a typical case) reduces to the critical lag for depinning studied in \cite{P11} and \cite{SP12} in the limit of no entrainment.
At the level of equation of motion for rigid lines, pinning could be modeled by introducing a local drag force with a non-constant $\eta$ 
parameter that depends on the dynamical lag $\omega$ as
\begin{align*}
    \eta(\bm x,\omega) \, \gg \, \kappa \rho_n(r)
    \qquad &\text{if } \quad |\omega(x)| \, <  \, \omega_{cr}(x)  \, \, ,
    \\
    \eta(\bm x,\omega) \, \ll \, \kappa \rho_n(r)
    \qquad &\text{if } \quad |\omega(x)| \, \simeq \, \omega_{cr}(x) \, \,.
\end{align*}
As an order of magnitude estimate, if $\eta \sim 10^{21}\,$s$\,$cm/g in the crust, 
the $\mathcal B^x$ of Eq \ref{eq:free-velocity} ranges from $\mathcal{B}^x(x=0) \sim 10^{-9}$
to $\mathcal{B}^x(x=R_d) \sim 10^{-12}$, thus ensuring that the vortices are expelled on timescales of many years.
\\
However, since we are just interested to estimate the angular momentum reservoir, we can rely to a simpler 
possibility that is more similar to the early vortex-creep model \cite{Alpar1984II}, where the effect of the dynamical lag $\omega$ against pinning 
is to modulate the  typical velocity of free vortices given in Eq \ref{eq:free-velocity}. 
This is equivalent to approximate the complete (and unknown) friction functional as
\[
    \mathcal{B}[\Omega_v,\Omega_p,x]\,\, \approx \,\,Y[\omega , x]\, \,\mathcal{B}^x(x) \, .
\]
From the  macroscopic point of view, the fraction $Y \in [0,1]$ can be equally interpreted as 
a parameter that slows the expulsion of vortex lines or the fraction of (unpinned) vortex lines that are free to move under 
the combined action of $\bm{f}_M$ and $\bm{f}_D$. By considering a large population of vortices,
$Y$ can also be seen as a local depinning probability. 
With macroscopically straight vortex lines, the simplest way to encode pinning at the macroscopic scale is thus to construct $Y$ in such a way that 
\begin{align*}
    Y[\omega , x] \, \approx 0 
    \qquad &\text{if } \quad |\omega(x)| \, <  \, \omega_{cr}(x)  \, \, ,
    \\
    Y[\omega , x] \, \approx 1
    \qquad &\text{if } \quad |\omega(x)| \, \simeq \, \omega_{cr}(x) \, \,.
\end{align*}
Finite temperature, quantum effects and finite vortex tension take part in smoothing the step-like form of $Y$.
As an explicit example, the original vortex-creep model \cite{Alpar1984II} in the regime $0 < \omega < \tilde{\omega}_{cr}$ is realized by
\[
    Y[\omega , x] \, \approx \, \exp \left(  \frac{\omega(x) - \tilde{\omega}_{cr}(x)}{\alpha \,\tilde{\omega}_{cr}(x)}  \right)  \, ,
\]
where $\alpha$ is a dimensionless temperature that tunes the local rigidity of the pinning  
and $\tilde{\omega}_{cr}(x)$ is an opportune critical lag. 
In the limit $\alpha \rightarrow 0$ the vortex-creep prescription results into a perfect pinning prescription, namely
$Y=\theta(\tilde{\omega}_{cr}(x) - |\omega(x)| )$ is the unit step function.
\\
\\
{\it Improving the friction term - } 
Vortices in an hydrodynamical model only serve to construct a particular form of the mutual friction in terms of the dynamical variables of the problem
(in our global model we shifted this problem on $\mathcal{B}$).
A glitch can then be triggered by perturbing $\mathcal{B}$ or, in order to 
mimic the sudden unpinning of many vortices, by suddenly increasing the $Y$ fraction in some interval of $x$.
It is thus important to find good prescriptions for the friction functional, even in global or averaged models like the one discussed here.
Other contributions to the vortex dynamics (that could have a significant impact on mutual friction) arise from the finite tension 
force of  vortices and their mutual interaction \cite{HM15,WS16}. 
Self-interaction forces that complicate further the picture are present for curved lines \cite{Qbook}.
Moreover the radiation of phonons during the pinning and depinning of a vortex line and proximity effect when two lines are close 
are important at a microscopic scale \cite{WM12}:
these precesses could provide the physics needed to explain the claimed self-organized criticality of glitching pulsars \cite{MP08}.
Further work on these lines is needed to understand how these issues modify mutual friction.
\section{Constraints from large glitches}
\vspace{2mm}
For any given lag $\omega$, the corresponding reservoir of angular momentum 
(i.e. the difference between the total angular momentum and that of the star in perfect corotation) is
\[
    L \,-\, I\,\Omega_p \,=\, I_v \, \langle \omega \rangle \, .
\]
The spin-up phase of glitches is so fast that the braking torque carries out a negligible amount of momentum; 
the total angular momentum is thus conserved during a glitch 
\[
    \dot{L} 
    \,=\, I_v \, \langle\dot{\omega} \rangle \,+\,  I \, \dot{\Omega}_p \,=\, 0 \, .
\]
Assume now to trigger a {\it maximal} glitch (a glitch that by definition empties the whole reservoir) at a certain initial time:
there will be a subsequent time at which the star is in a state of instantaneous average corotation, namely $\langle \omega \rangle =0$.
This situation can be realized during the spin up phase or just after an overshoot, a transient period during which $\langle \omega \rangle<0$. 
Integration of the previous equation between these two times gives
\[
    I_v \, \langle \Delta \omega \rangle \, + \, I \Delta \Omega_p \, = \, 0 \, ,
\]
where $\Delta$ indicates the difference between a quantity at the particular time when corotation is instantaneously realized and the same quantity 
at the triggering time.
\\
\\
{\it Entrainment independent maximum glitch amplitude - }
In our model the maximum sustainable lag is the critical depinning lag $\omega_{cr}$ 
(the actual maximum sustainable lag is likely to be less than this quantity, because of thermal effects and dynamical creep)
so that we can set an absolute upper limit to the glitch amplitude by imposing $\Delta \omega \, = \,- \omega_{cr}$:
\begin{equation}
\label{eq:max-gl}
    \Delta \Omega_{max} \, = \, \frac{I_v}{I} \, \langle\omega_{cr} \rangle \,.
\end{equation}
Using the explicit forms of $dI_v$ and $\omega_{cr}$, it is trivial to show that $\Delta \Omega_{max}$ depends only 
on the EOS and on the pinning forces \cite{AP16}: at the Newtonian level entrainment does not affect this quantity 
(this ceases to be the case when relativistic corrections are taken into account). 
Thus, by fixing these two microphysical inputs, it is possible to study how $\Delta \Omega_{max}$ changes with the total mass of the NS. 
This is shown in Fig 3, together with the maximum glitch amplitude $\Delta\Omega$ recorded in some pulsars;
we thus find an upper limit to the mass of each pulsar by imposing $\Delta\Omega=\Delta\Omega_{max}(M)$.
As a general trend, a stiff EOS like GM1 gives less stringent upper mass bounds (e.g. $M_{Vela}<1.4$ with SLy and 
$M_{Vela}<1.8\, M_{\odot}$ with GM1). Future observations of even larger glitch amplitudes can only lower this upper limit to the mass, leading to stronger constraints. 
Note also that the possible presence of turbulence in the crust could reduce the maximum glitch amplitude, thus lowering the curves of Fig 3 and leading to stronger constraints: 
it seems reasonable to expect that the increase in the vortex length density due to the development of complex vortex tangles weakens the otherwise coherent pinning of the vortex array. Finally we remark that, as long as only crustal pinning is considered, $\Delta \Omega_{max}$ 
turns out to be the same for vortex lines that stop at the crust-core interface or pass through the core.
%
%
%
%
\begin{figure}
\includegraphics[width=0.96\textwidth]{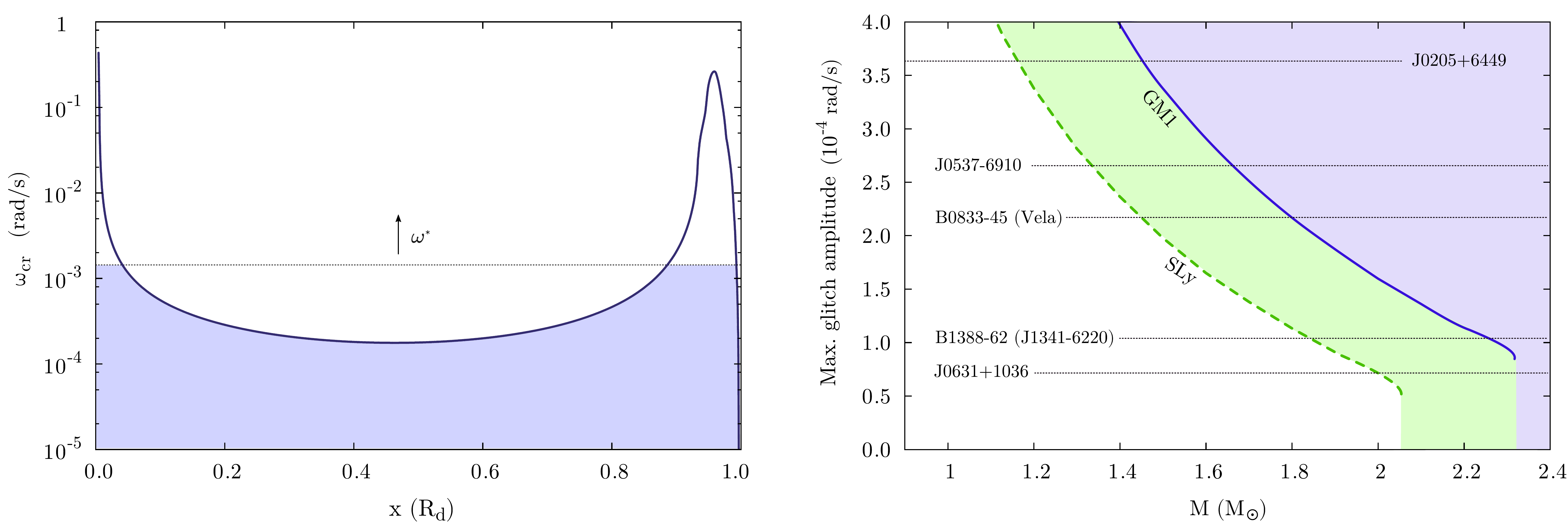}
\raggedright
\begin{minipage}[t]{17.5pc}
       \label{fig:wcr}
       \caption{Example of the critical lag for unpinning $\omega_{cr}$ for a NS of mass $1.1\,M_{\odot}$ and the SLy equation of state.
           The horizontal line indicates the increasing nominal lag $\omega^*$,  
           the shaded area below it is bounded by the corresponding lag developed since corotation. 
           The distance from the rotational axis is expressed in units of the neutron drip radius $R_d$.
           \\
        }
\end{minipage}
\hspace{1.5pc}
\begin{minipage}[t]{17.5pc}
       \label{fig:max-gl}
       \caption{
           The theoretical maximum glitch $\Delta\Omega_{max}$ is shown as a function of the stellar mass $M$
           for the Sly equation of state (green, dashed) and the GM1 (blue, solid). 
           The curves terminate at the maximum mass allowed by each EOS.
           Horizontal lines indicate the largest glitch $\Delta \Omega$ recorded in the
           corresponding pulsar. The shaded area defines the forbidden region.
           \\
        }
\end{minipage}
\end{figure}
%
%
%
%
\begin{figure}[h]
    \raggedright
    \begin{minipage}[c]{22pc}
    \includegraphics[width=23.5pc]{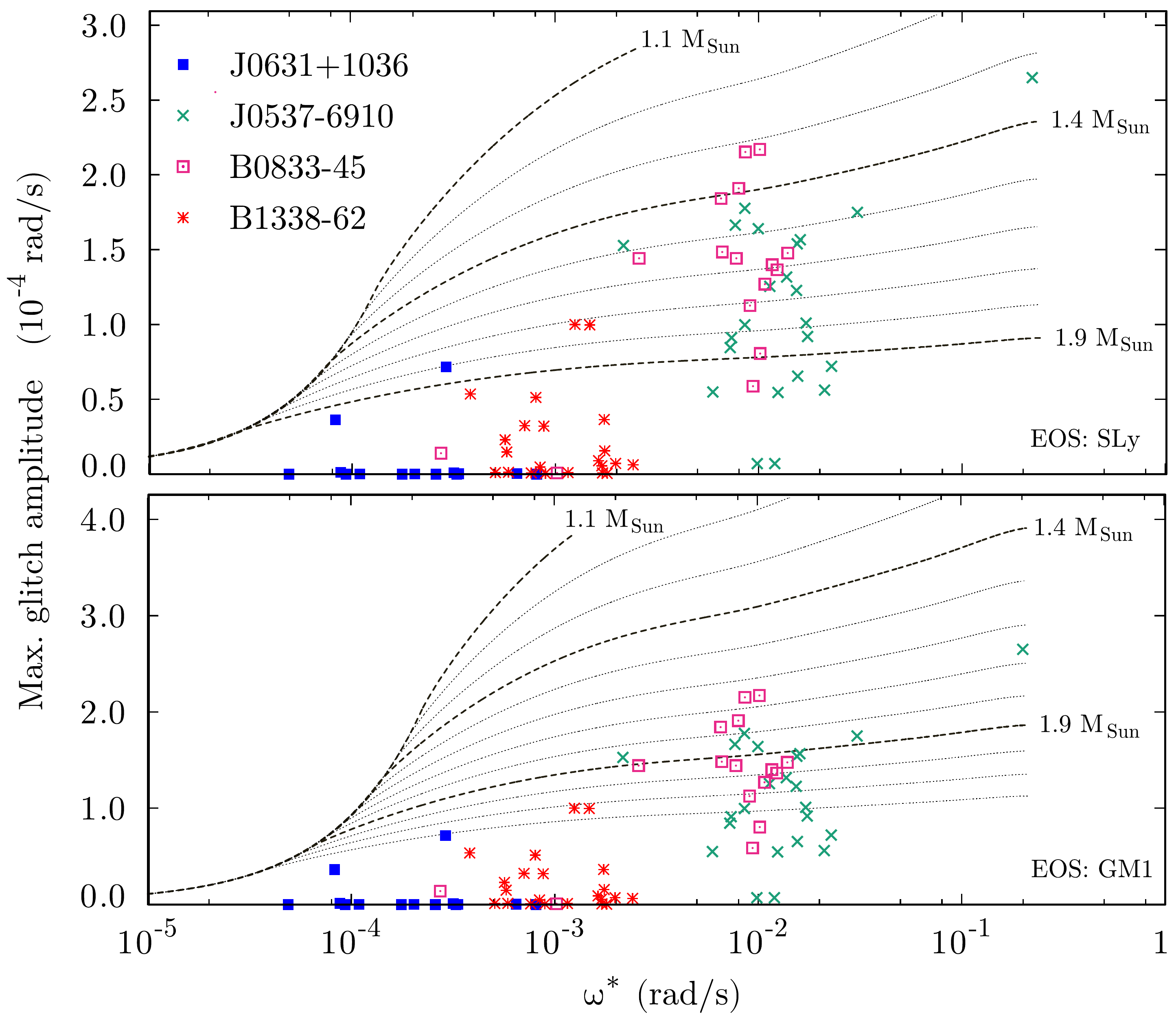}
    \end{minipage}
    \hspace{2pc} 
    \begin{minipage}[c]{13pc}
    \label{fig:lines-sly-gm1}
    \caption{Upper panel: the lines defined in Eq \ref{eq:unified-model} for different masses relative to the SLy equation of state 
        are shown as a function of the nominal lag $\omega^*$.
        For comparison we superimpose the single glitches of four different pulsars; the nominal lag of each glitch is given by $-\dot \Omega t$,
        with $t$ the waiting time since the previous event and $\dot\Omega$ the spin down rate measured in the corresponding pulsar.
        Lower panel: the same but using the GM1 equation of state.
        \\
    }
    \end{minipage}
\end{figure}
\\
\\
{\it Constraints imposed by the reservoir dynamics - }
It is possible to try to lower the upper bound of the mass by using also observed timing properties other than the largest glitch.
Starting from corotation, forward integration in time of Eqs \ref{eq:1} and \ref{eq:2} allows to follow the evolution of the quantity
\begin{equation}
    \label{eq:dynamical-max-gl}
    \Delta\Omega_{m}(t) \, = \, \frac{I_v}{I} \, \langle \omega(x,t) \rangle \,
\end{equation}
Equation \ref{eq:dynamical-max-gl} gives the amplitude of a maximal glitch, triggered at time $t$ after the emptying of the reservoir because of an initial large event.
This results in a correlation between the amplitude of maximal glitches and the waiting time between them.
This of course ceases to be interesting if there are no pulsars that sporadically empty the reservoir: in this case the sequence 
of amplitudes and waiting times shows no correlations between events, which is certainly true for glitches of small amplitude that comprise the major part of the whole sample 
among the pulsar population. 
However J0537-6910 and Vela have been proven to be exceptional objects:
systematic study of glitches in individual pulsars seem to indicate that the distribution of waiting times is Poissonian while 
the distribution of glitch amplitudes is a power law (even though the value of the exponent is not universal across the pulsar population), 
but Vela and J0537-6910 showed evidence for quasi-periodicity \cite{MP08}.
The main difficulty behind this and similar analysis is the paucity of data in single objects.
\\
In the following we present a more speculative set of ideas: instead of solving dynamical equations 
with the particular rotational parameters for each pulsar, we use the simple unified model sketched in Fig 2
and defined as
\begin{equation}
\label{eq:unified-model}
\Delta\Omega_{m}(\omega^*) \, = \,  \frac{I_v}{I} \, \langle \, \min{(\omega_{cr}(x),\omega^*)}   \, \rangle \,.
\end{equation}
For each pulsar we measure time in terms of a nominal lag $\omega^* = t\, \dot\Omega_{\infty}$.
The curves defined by Eq \ref{eq:unified-model} are shown in Fig 4 for different masses.
On top of this, we plot the single glitches of some pulsars (among which Vela and J0537-6910) versus waiting times to the previous glitch 
multiplied by the absolute value of the spin-down rate $-\dot\Omega$ the object.
The largest glitch of J0537-6910 (the only pulsar that display correlation between the amplitude and the time to the next glitch \cite{HM15})
is also the first one, so we plotted it at the end of the $\omega^*$ axis, where $\Delta\Omega_{m}(\omega^*)$ assumes the value given by Eq \ref{eq:max-gl}.
It is interesting to note that  is always the largest glitch that touches the curve corresponding to the smaller mass, even if the data are scattered. 
Note that for each glitch we used the nominal lag corresponding to the time since the previous event, without taking into account the
amplitude of this event, even if it is unlikely that small glitches are maximal. 
This only serves as a clue of the fact that the largest glitch recorded in a pulsar can constrain its mass:
at this level no conclusions can be taken  on the actual (possibly time-correlated) occurrence of large glitches in pulsars. 
\\
\\
{\it A remark on the unified model - }
 We stress that when we simulate the full dynamical equations, starting from perfect corotation at $t=0$,
 the lag evolves in a fashion very similar to the lag sketched in Fig 2: the horizontal plateau (at height $\omega^*$) rises
 until it locally approaches $\omega_{cr}(x)$, where sub-critical vortex creep shapes the actual lag around $\omega_{cr}$.
 Thus it is not a bad approximation to recast the dynamical evolution of the lag into the form 
 \[
 \omega(x,t) \, \approx \, \min{\left(\, \omega_{cr}(x) \, , \, \omega^*(t)\,\right)} \, .
 \]   
 However this can only serve to {\it define} $\omega^*(t)$, a function of time that can be easily extrapolated from
 $\omega(x,t)$ once the actual evolution is known from the forward integration in time of the dynamical equations
 (e.g. $\omega^*(t)=\omega(x_m,t)$ where $x_m$ is the cylindrical radius corresponding to the peak of $\omega_{cr}$).
 On the other hand, in the previous section we assumed $\omega^*(t) = \dot\Omega_{\infty}t$.
 This choice is convenient since it is possible treat different pulsars within a unified prescription; in this way there is no need to solve the dynamical 
 equations with the specific rotational parameters $\Omega_p(0)=\Omega$ and $\dot \Omega_{\infty} = -\dot \Omega$ for every different object.
 However there are no rigorous arguments for doing this, except for the fact that the slow dynamics 
 of $\omega$ toward the steady state is driven by $\dot \Omega_{\infty}$. Moreover this unified prescription based on a linear relation between the actual time and the 
 nominal lag $\omega^*$ was already used (in a different context and with a different idealization of the lag dynamics) 
 in some previous works \cite{P11,SP12,HP13} that use the same form of the critical profile given in Eq \ref{eq:clag}, but without entrainment.
\begin{figure}[h]
    \centering
    \includegraphics[width=.9 \textwidth]{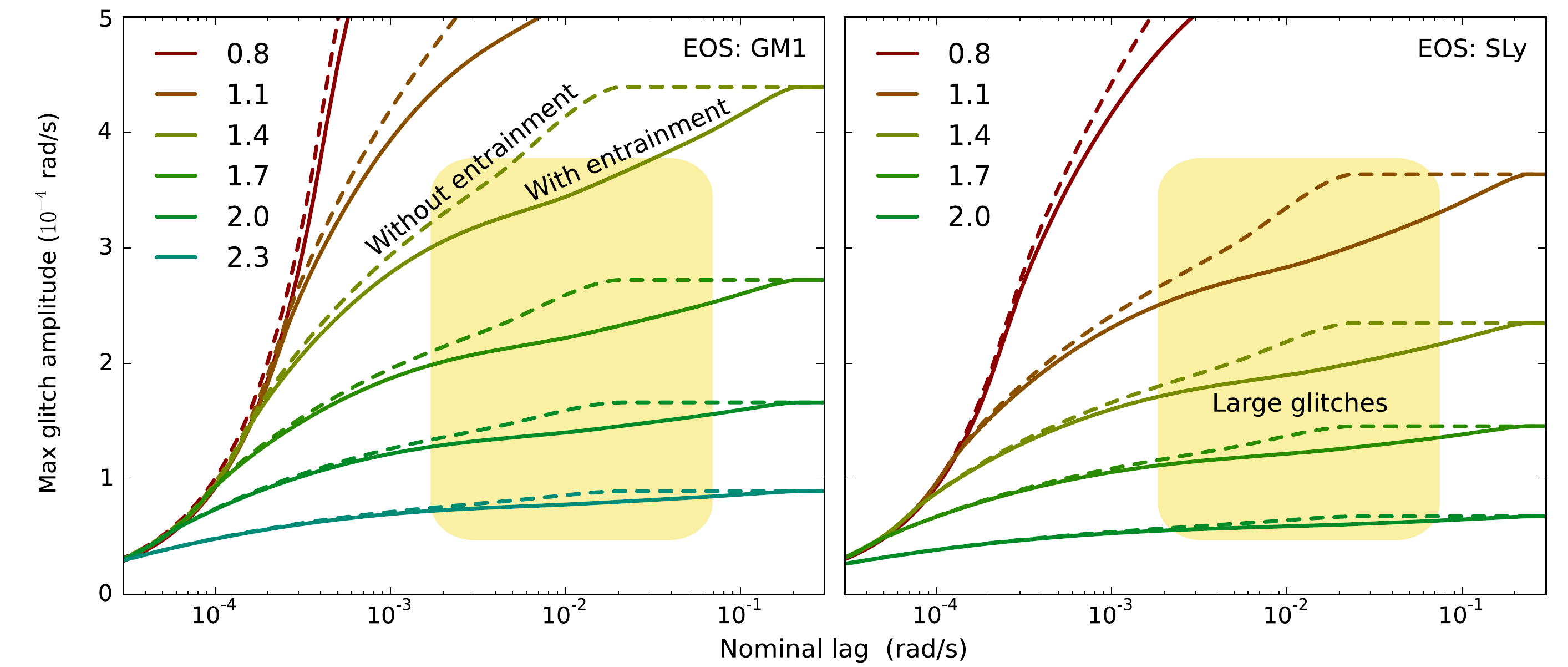}
    \caption{Right: comparison between the maximal glitch amplitudes given by Eq \ref{eq:unified-model} when entrainment effect is taken into account and when $\epsilon_n=0$ everywhere.
        The EOS used is GM1 and the different curves refer to the different mass values (in units of $M_{\odot}$) listed in the key. 
        The region where large glitches are expected is shaded: it extends circa to the value
        of the largest glitch recorded in J0205-6449. The $\omega^*$ scale on the horizontal axis spans the range $3\times10^{-5}\div0.3\,$rad/s.
        Right: the same but using SLy equation of state.
    }
    \label{fig:entr-noentr}
\end{figure}
\section{Effect of entrainment}
\vspace{2mm}
We switch off the entrainment effect setting $m^*(r)=1$ everywhere and repeat the integrations needed to calculate the  macroscopic input quantities of model.
In this limit $\omega \rightarrow \Omega_n-\Omega_p$, $I_v \rightarrow I_n$, $I_p \rightarrow I-I_n$.
%
The curves $\Delta\Omega_{m}(\omega^*)$ of the unified model (Eq \ref{eq:unified-model})  are found 
to be considerably steeper than their ``entrained'' counterparts, as shown in Fig 5.
This is not surprising since the strong crustal entrainment sensibly increase the peak of the critical lag which in turn implies 
that the plateau  at the value $\Delta\Omega_{max}$ is reached for smaller values of the nominal lag.
This is in part  a drawback of the assumed form of $\omega^*(t)$ that in general is not simply linear in time. 
Nonetheless the result seems physically reasonable as it is more difficult 
for strongly entrained superfluid neutrons to lag behind the crust \cite{AG12}. However a true comparative study of entrained and non-entrained 
dynamics should be carried by solving the dynamical equations \ref{eq:1} and \ref{eq:2}.
Despite this caveat this preliminary analysis is encouraging: large glitches of Vela and J0537-6910 seem to have been triggered around 
$\omega^* \sim 10^{-2}\,$rad/s, as can be seen in Fig 4.
For this value of the nominal lag the difference between the case with entrainment and that without entrainment results in at most 
a $\sim 10\%$ uncertainty on the mass value.
%
%
%
\section{Conclusions}
\vspace{2mm}
We presented a global model for pulsar rotational dynamics that accounts consistently for the layered structure of the star, 
the differential rotation of the superfluid and the presence of density-dependent entrainment \cite{AP16}. 
The Eqs \ref{eq:1} and \ref{eq:2}
describe the exchange of angular momentum between two effective components fo the star, 
characterized by the partial moments of inertia $I_v$ and $I_p$.
All the complex behavior of the rearrangement of  the vortex configuration that should appear in Eq  \ref{eq:2}
is hidden into the functional $\mathcal{B}$.
%
Even without solving explicitly the dynamical equations, it is possible to draw some general quantitative conclusions; in particular 
the calculation of the maximum glitch amplitude (that is entrainment independent and is 
unaffected by the possible presence of a normal layer at the crust-core interface) seems to open two possibilities:
\begin{enumerate}
\item[-] Test pinning newly calculated forces using observations of large glitches.
\item[-] Set an upper limit to the mass of some isolated pulsars (once the EOS and the pinning force profile have been fixed).
\end{enumerate}
Finally we sketched a more speculative set of ideas that could be useful to refine the upper bound to the mass of large glitchers: while the absolute upper limit 
is fixed by the observation of the largest glitch only, to improve this constraint we have to use some additional information about the temporal occurrence of glitches 
in a pulsar.
\\
A final remark is due: we used the TOV equations for the structure of the star, but the classical moments of inertia.
Fortunately it is quite simple to introduce general relativistic corrections for slow rotation to $I$ and $I_v$. 
Preliminary study indicates that (when vortex lines extend into the core, as in this presentation) the quantity $I_v/I$ is nearly unaffected by relativistic corrections.
However general relativity does not affect only the moments of inertia but also the measure $dI_v$ and 
the critical lag; since in this Newtonian model we used the paraxial vortex configuration of a rigidly rotating body, the corrections to the cylindrical profiles could be estimated by resolving the stationary configuration of vortex lines for a slowly and rigidly rotating neutron star.
Despite all possible caveats and simplifications that are present in our model, the message we want to vehicle is that 
also the study of single glitches (and in particular the largest glitch recorded in each pulsar) can robustly constrain 
the mass of isolated neutron stars: large glitches can provide a test for the consistency of glitch theory that is alternative 
to the studies based on pulsar activity \cite{LE99,AG12,C13}.
\section*{Acknowledgments}
\vspace{2mm}
This work consists of a revised version of the presentation given by MA at the
{\it{NewCompStar Working Group 2 Meeting}}
, 23-27 May 2016, GSSI and LNGS (L'Aquila, Italy). 
Partial support comes from NewCompStar, COST Action MP1304.
\\
\\
MA wishes to express his gratitude to Stella Valentina Paronuzzi Ticco
(previously at International School for Advanced Studies in Trieste) for valuable help and thoughtful discussion
and also to thank Nicolas Chamel for interesting comments on some aspects of entrainment.
%
%
\section*{References}
\vspace{2mm}
%
%
\providecommand{\newblock}{}

\end{document}